\documentclass[conference]{IEEEtran}
\IEEEoverridecommandlockouts
\usepackage{amsmath,amssymb,amsfonts}
\usepackage{algorithmic}
\usepackage{booktabs}
\usepackage{graphicx}
\usepackage{textcomp}
\usepackage{xcolor}
\usepackage{xspace}
\usepackage{balance}
\usepackage[style=ieee]{biblatex}
\def\BibTeX{{\rm B\kern-.05em{\sc i\kern-.025em b}\kern-.08em
    T\kern-.1667em\lower.7ex\hbox{E}\kern-.125emX}}
\begin{document}

\title{Towards Autonomous Testing Agents via Conversational Large Language Models %\\
}
%\title{\name~- Towards Autonomous Testing Agents via Conversational Large Language Models %\\
% \thanks{Identify funding agencies here or move to ack.}
%}

\author{\IEEEauthorblockN{Robert Feldt}
\IEEEauthorblockA{
\textit{Chalmers University of Technology}\\
robert.feldt@chalmers.se}
\and
\IEEEauthorblockN{Sungmin Kang}
\IEEEauthorblockA{
\textit{KAIST}\\
sungmin.kang@kaist.ac.kr}
\and
\IEEEauthorblockN{Juyeon Yoon}
\IEEEauthorblockA{
\textit{KAIST}\\
juyeon.yoon@kaist.ac.kr}
\and
\IEEEauthorblockN{Shin Yoo}
\IEEEauthorblockA{
\textit{KAIST}\\
shin.yoo@kaist.ac.kr}
\and
}

\newcommand{\cell}[1]{\begin{tabular}{@{}c@{}}#1\end{tabular}}
\newcommand{\name}{\textsc{SocraTest}\xspace}

\newcommand{\rf}[2][red]{\textcolor{#1}{RF: #2}}
\newcommand{\sm}[2][green]{\textcolor{#1}{SK: #2}}
\newcommand{\sy}[2][blue]{\textcolor{#1}{SY: #2}}
\newcommand{\jy}[2][teal]{\textcolor{#1}{JY: #2}}
\newcommand{\fixme}[2][red]{\textcolor{#1}{FIXME: #2}}
\newcommand{\justred}[2][red]{\textcolor{#1}{#2}}
\newcommand{\addcite}[2][orange]{\textcolor{#1}{ADDCITE: #2}}

\maketitle

\begin{abstract}
Software testing is an important part of the development cycle, yet it requires specialized expertise and substantial developer effort to adequately test software. Recent discoveries of the capabilities of large language models (LLMs) suggest that they can be used as automated testing assistants, and thus provide helpful information and even drive the testing process. To highlight the potential of this technology, we present a taxonomy of LLM-based testing agents based on their level of autonomy, and describe how a greater level of autonomy can benefit developers in practice. An example use of LLMs as a testing assistant is provided to demonstrate how a conversational framework for testing can help developers. This also highlights how the often criticized ``hallucination'' of LLMs can be beneficial for testing. We identify other tangible benefits that LLM-driven testing agents can bestow, and also discuss
potential limitations.
%of them as well.
\end{abstract}


\section{Introduction}

% below is my attempt at an introduction - sungmin

% \fixme{I don't particularly like this starting paragraph, any suggestions welcome}
Software testing, an integral part of the development cycle, enables quality assurance and bug detection prior to deployment, for example via continuous integration practices~\cite{Shanin2017CI}.
%Software testing is an important part of the development process, as bugs can be detected before the code is deployed, for example via continuous integration~\cite{Shanin2017CI}. 
%However, (automated) software testing can be difficult, and require technical familiarity to set up. 
However, automated software testing can be challenging, and necessitates a high level of technical acumen. 
There is significant expertise required to appropriately test software, as evidenced by the existence of test engineers/architects. Meanwhile, as Arcuri~\cite{Arcuri2018ExperienceRep} notes, existing automated software test generation tools may also require significant expertise in the tool, in addition to being difficult to apply in industry.% due to factors such as user interface.
Furthermore, software testing and the writing of software tests can be repetitive, as Hass et al.~\cite{Haas2021ManualTesting} note. A more positive attribute of test cases is that their syntax is often significantly simpler than %when compared to 
production software~\cite{Rutherford2003TestComplexity}.
%Regarding setting up the environment, mostly the same procedure can be used when using the same testing framework, and the syntax of tests is significantly simpler when compared to production software~\cite{Rutherford2003TestComplexity}. 

These unique characteristics of test code naturally bring about a distinction between testing experts and domain experts, which existing literature on developer expertise~\cite{liang2023qualitative} supports by identifying distinct types of expertise: ``understanding the vision of the project'' and ``knowledge about tools''. Under this framework, an ideal setup would be one in which
a testing expert and a domain expert collaborate to write tests for a project. The domain expert may lay out the specifications of a project, while the testing expert may convert those specifications into concrete tests, based on the testing expert's experience. A great strength of this process is that as a result of such a dialogue, initially unexpected, yet nuanced issues with the specification may arise, which provide opportunities to clarify the desired behavior. Indeed, handling such unexpected behavior is one of the virtues of software testing~~\cite{Feldt:2016if,Chen:2010kq}.

In this paper, we argue that Large Language Models (LLMs), which have been trained with a large quantity of code including software test data~\cite{gao2020pile} may eventually be capable of providing such testing knowledge, and that humans may act as domain experts and specify or clarify to the LLM what the intended behavior is.
%who can simply specify what the intended behavior is to the LLM. 
Specifically, we argue that LLMs are sufficiently well-trained with software tests to `fill in' lower-level details of the intention of the developer.
They also exhibit some `knowledge' about testing methodologies,and can adapt them to new situations~\cite{jalil2023chatgpt}.
% too short
Going further, LLMs appear sufficiently capable in dialogue to converse about the results with a prospective software tester % need to unify these terms at somepoint
so that they could engage in a `Socratic' manner: that is, they could provide counterexamples to help the developer to think their specification through, and thus uncover unexpected issues with the desired behavior, in this process clarifying what would be ideal. Equipped with appropriate `middleware' which provides tools that the LLM could interact with, our eventual vision is that we can grant the LLM `autonomy', in which it would set up a plan and `use' the tools at its disposal to achieve the high-level objective set by the developer while abstracting away lower-level details.

To illustrate this idea, we organize the paper as follows. In Section~\ref{sec:background}, we present literature on LLMs and how they can be used to emulate cognitive models for human behavior, thus providing a way of implementing our vision of testing LLMs that interact with tools and have agency while interacting with humans. In Section~\ref{sec:vision}, we provide a taxonomy of LLM-based software testing systems based on whether the LLMs are used in an \textit{interactive} way, and the degree of `\textit{autonomy}', i.e. formulating and executing its own plans. 
%to what degree the LLM-using tools are `autonomous', in the sense of whether the tool can formulate and execute a plan that the tool itself came up with. 
In Section~\ref{sec:examples} we present an example interaction with the GPT-4 model~\cite{openai2023gpt4}, demonstrating that even without significant autonomy, developers gain an opportunity to ponder %about the 
fine-grained semantics of their code via dialogue. The benefits of (autonomous) conversational testing agents are given in Section~\ref{sec:progress2v}, and we argue that greater autonomy confers greater benefits. Potential limitations are given in Section~\ref{sec:limitations}, and conclude in Section~\ref{sec:conclusion}.

\section{Background}
\label{sec:background}
% AutoGPT (current movement) -> classical models -> village paper
 
With the recent advancements in large language models, the possibility of having a personal agent capable of assisting with general tasks is greater than ever. One popular ongoing project is AutoGPT~\cite{autogpt2023}, which aims to implement a fully autonomous system that works towards achieving user-defined goals. Beyond the basic capabilities of GPT-4 model, the proposed framework supports a high level of autonomy by giving access to external tools such as search engines, complementary neural network models, and file systems. Moreover, AutoGPT retains both short-term and long-term memory management to cope with complex tasks that exceed the context length limitation of currently available language models. 

The operation of AutoGPT can be interpreted from the perspective of existing cognitive architecture frameworks. In fact, modelling the human mind has been a longstanding research interest, driven by both objectives of explaining human behaviors and devising artificial intelligent agents. Several influential architectures such as ACT-R~\cite{anderson2009can}, and SOAR~\cite{laird2019soar} have been developed so far, and their core components contain associative memory structures linked with perception and actuation (``motor'')~\cite{laird2022analysis}. This bears resemblance with AutoGPT's architecture: i.e., incorporating external tools to perceive new information (e.g., via search engine results) or perform an action (e.g., writing a Python script) may be viewed as building the perception and actuation modules into the architecture. On the other hand, LLMs can strengthen classical cognitive architectures by deriving plausible actions using the relevant memory and current state as prompting context.

Park et al.~\cite{park2023generative} provided an interesting case study, where a memory-enabled framework that embeds LLMs with a unique reflection strategy on stored memories is used to simulate multi-agent social interactions. By prompting a LLM, their agent architecture continuously induces higher-level interpretation on what the agent had perceived. This enables the agent to maintain long-term coherence of its own behaviour, and in that process, plausible emergent social behavior is simulated. The recent paper by Wang et al.~\cite{wang2023voyager} also shows an LLM-based architecture that can explore a 3D world, acquire diverse skills, and make novel discoveries without human guidance.

Such advances in developing cognitive architectures on top of LLMs also open up numerous possibilities for software testing automation, such as managing continuous testing history as memories and planning the general testing strategy and then trying to fulfill sub-goals of the plan. In addition, an autonomous testing agent could evolve the test suite on its own by allowing the architecture to execute existing testing tools and access the results. In the following section, we provide a series of steps towards implementing such a testing agent.

\section{Vision - \name}
\label{sec:vision}

Based on existing research on LLMs, our vision is to build \name, a framework for conversational testing agents that are potentially autonomous and supported by existing automated testing techniques via a dedicated \emph{middleware}, that would invoke appropriate tools based on LLM output, so that LLMs can operate in an autonomous manner. We posit that such an agent can not only become an intelligent testing partner to a human software engineer, but also be able to handle typical testing related tasks autonomously. 

A taxonomy of LLM use when performing software testing is presented in Table~\ref{tab:taxonomy}, with higher rows representing higher degrees of autonomy from the LLM perspective. Specifically, the Driver column shows who drives the operation, i.e., who initiates the task, collects information, and decides the next step. For example, code completion provided by GitHub Copilot is automatically initiated by the front-end, i.e., the editor. Techniques based on contextual prompting, such as Libro~\cite{kang2022large}, are still considered to be driven by the technique itself, in that a human is the starting point but not part of the workflow. Conversational testing, an example of which is shown in Section~\ref{sec:examples}, involves a human in the interactive loop: the user drives the next action via the dialogue with the LLM. 

We can also categorize LLM usages based on their information sources: more advanced use cases increasingly involve a wider range of information sources and more complicated interactions. In the most basic usage of LLMs, i.e. auto-completion and in-filling, the only information source is the code context, which is already written by the human user. In contrast, Contextual Prompting provides further contextual information, e.g. in the form of examples, and depends on the few-shot learning capabilities of LLMs to perform the given task. While this approach has successfully enabled much more complicated tasks such as bug reproduction~\cite{kang2022large}, its format is still a single query-and-response, without any interaction between the developer and the tool.

We argue that a tool capable of dialogue, corresponding to Conversational Testing and upward in the taxonomy, can extend the scope of both the role of the driver and the information sources and enable unique benefits (as in Section~\ref{sec:examples}). At the lowest level of autonomy (Conversational Testing), as a conversational partner, LLMs partially drive the process, but only respond to human requests without autonomy. One level up, we can introduce a low level of autonomy by providing codified instructions for the LLM to follow (Conversational Testing with Tools): for example, we can set structural testing as a goal and allow LLMs to initiate the use of appropriate tools, e.g. EvoSuite~\cite{Fraser2010Evosuite} and Jacoco~\cite{jacoco2013}, to generate tests and measure coverage. Finally, at the highest level of autonomy (corresponding to Conversational Testing Agents), LLMs are augmented with memory and planning capabilities so that humans only need to provide high-level directions, while LLMs initiate and complete whole tasks of a testing process.

To implement such autonomous testing agents using LLMs, a prerequisite is the implementation of middleware for conversational testing agents as a set of supporting features. Various existing testing tools and techniques should be included in the middleware so that they can be used by the LLM. The middleware can also augment LLMs with memory, similarly to experiments such as AutoGPT~\cite{autogpt2023} or other autonomous cognitive models based on LLMs~\cite{park2023generative}. This middleware may use frameworks such as LangChain~\cite{chase2022LangChain}, which ease the connection between LLMs and external tools.
% While demonstrating the capability of such fully autonomous testing agents would be ideal, it would require a significant implementation effort. 
In lieu of  the fully realized vision, we present how even at a lower level of autonomy, i.e. at the conversational testing level, testing can become much easier from the developer's perspective.

\begin{table*}
\caption{Taxonomy of LLM Uses in Software Testing\label{tab:taxonomy}}
\begin{tabular}{lcclc}
\toprule
Mode of Usage & Driver & Interactive & Available Information & Autonomy \\ \midrule
Conversational Testing Agents & \cell{Human,\\Middleware,\\LLM} & Yes & \cell{Extensive:, information from both user and the tools in middleware} & High \\ \midrule
Conversational Testing with Tools & \cell{Human,\\Middleware} & Yes & \cell{High, additional outputs from algorithms \& methods} & Low \\ \midrule
Conversational Testing & \cell{Human} & Yes & \cell{Rich: a mixture of templates, contexts, examples, and explanations} & No \\ \midrule
Contextual Prompting & \cell{Front-end,\\Testing SW} & \cell{No} & \cell{Medium: templates with contexts \& examples} & No \\ \midrule
Completion \& Infilling & \cell{Front-end,\\Testing SW} & No & \cell{Low: typically autocompletion of given code} & No\textbf{} \\
\bottomrule
\end{tabular}
\end{table*}

\section{Inspirational example tasks}
\label{sec:examples}

We have had a large number of software testing related conversational interactions with the GPT-4 model through the ChatGPT interface. We have found that the model can both describe different types of testing methods, merge and condense them to checklists to guide testers, as well as write executable test code to apply and exemplify the methods\slash checklists. We have also found the conversational mode essential both to clarify, as a developer or tester, the type of testing and support one needs and to request additional test code and the use of additional testing methods. For brevity, we here provide only a condensed example of a multi-step interaction we had
% \footnote{The Xth author had the interaction with GPT-4 in preparation for, and also described it on a high level in, his NEXTA 2023 Keynote in Dublin, Ireland on April 20th 2023} 
with the model to do unit testing for the Julia language~\cite{bezanson2012julia}, with each user query marked with "Prompt N".

After describing the type of Julia code we developed we asked GPT-4 for concrete advice, methods and checklists for how we should write unit tests (Prompt 1). It provided a detailed and long checklist that gave general and broad advice. It was actionable but quite generic. We then asked it to focus on test input selection and to provide a more detailed method and checklist (Prompt 2). GPT-4 proposed that we should use ``Equivalence Partitioning'' and ``Boundary Value Analysis'' and went on to define them. It also proposed a checklist that combined the main steps of the two techniques. We then asked it to provide example Julia test code to test a function in Julia's Base library that takes 2-3 inputs (Prompt 3). The model selected the \verb!Base.clamp(x, lo, hi)! function and briefly described it (``The clamp function restricts a value x to be within the range [lo, hi]. If x is less than lo, it returns lo. If x is greater than hi, it returns hi. Otherwise, it returns x.''). It then provided Julia test code with 16 test cases, an excerpt of which is shown below. It grouped test cases in relation to its checklist and briefly documented each group to indicate the checklist item that ``leads'' to the group.

% \begin{figure}[ht]
\begin{small}
\begin{verbatim}
using Test

@testset "clamp function tests" begin
    # Valid input classes
    # Typical inputs
    @test clamp(5, 2, 7) == 5
    @test clamp(1, 0, 10) == 1
    @test clamp(6, 0, 10) == 6
    
    # Boundary cases
    @test clamp(0, 0, 10) == 0
    @test clamp(10, 0, 10) == 10
    @test clamp(5, 5, 5) == 5
    
    # Invalid input classes
    # lo > hi
    @test_throws ArgumentError clamp(5, 7, 3)
    <...9 test cases omitted for brevity>
end
\end{verbatim}
\end{small}
% \caption{The initial test cases for clamp generated by ChatGPT\label{fig:test1}}
% \end{figure}

Based on our limited experience of the \verb!clamp! function we did not spot any problems with the test code; all test cases initially made sense. Upon execution, it turned out that 15 test cases pass but that an \verb!ArgumentError! is \textbf{not} thrown by Julia when calling \verb!clamp(5, 7, 3)!; Julia actually returns 3. We told the model about this (Prompt 4) by providing the output from running the test code as well as from making the call and showing the value returned by Julia. We asked it to explain why this happened. The model apologized and explained that Julia actually returns the \verb!hi! value in cases where it is lower than the \verb!lo! value. It went on to update the test code and corrected the \verb!@test_throws ArgumentError ...! as shown in the following.

% \begin{figure}[ht]
\begin{small}
\begin{verbatim}
# Invalid input classes
# Julia return hi if lo > hi
@test clamp(5, 7, 3) == 3
\end{verbatim}
\end{small}
% \caption{Failure from Generated Tests\label{fig:result}}
% \end{figure}

We were not impressed by the fact that the model were now confidently explaining that the \verb!clamp! function behaves in this way when it had earlier proposed this was not the case. However, the conversational mode of interaction was useful in nudging the model to give us more detailed and concrete information and in particular to provide relevant test code to exemplify its recommendations. It seems clear that this can have pedagogical and learning benefits as well as act as a reminder to apply important testing techniques in new contexts. The interactive, conversational mode also allowed us to further explain what we meant and requested and to ask the model to update and refine test code it had earlier provided.

We also argue that the ``erroroneous'' test code provided for the \verb!lo > hi! case shows that LLMs like GPT-4 can be particularly useful for testing. While the ``error'' the model did in the earlier step can be seen as a type of hallucination~\cite{ji2023survey}, we argue that for testing this is less severe (test code will not be part of the final, deployed software system) and can even be a benefit. In this case we argue that even a human tester could have assumed that the clamp function would first ensure that the lo value is less than or equal to the hi value and that an exception would be thrown otherwise. We actually learnt something about Julia through this mistake and we argue that a tester and developer could also have learnt something and even decided that raising an exception would be the more sensible thing to implement. In this sense, for software testing, the so called ``hallucination'' that LLMs have been criticized for can, at least sometimes, be a benefit, as it can prompt deeper thought. This is in line with the argument of Feldt et al.~\cite{feldt2018ways} that ``process'' use of AI in software development is less risky.

\section{Progress towards vision}
\label{sec:progress2v}

% \rf{Highlight at least 3 benefits we see: hallucination can be a good thing in testing (even if not in general or when coding), can implement also non-formalized testing methods (like scientific debugging paper) by describing them in natural language, can combine both existing, exact methods (static analysis, code coverage etc) and make high-level plans... some other benefits we might want to touch on: also it can adapt general testing ideas to a specific case, can help democratize testing since same system can simply be used also by less experienced testers and companies}

% - Codifying patterns of LLM interaction / prompt engineering
% - Adding test specific knowledge (metrics, testing goals,...)
% - Create general middleware 

While even low-autonomy conversational testing can help the developer verify software, techniques with higher autonomy can confer even greater benefits. We identify that there are at least three benefits to conversational testing via LLMs, which are increasingly "unlocked" with a higher level of autonomy. To start, as mentioned earlier, while LLM hallucination has been identified as a problem~\cite{bang2023multitask}, it can actually be an asset when doing software testing, as in general we want to be able to generate tests that uncover the unexpected behavior of software~\cite{Feldt:2016if,Chen:2010kq}. This characteristic benefits all levels of LLM use for testing, as `hallucination' can happen at any level of content generation while using LLMs. 

At a greater level of autonomy (Conversational Testing with Tools or higher), we argue that one of the greatest benefits LLMs can bring about is the fact that they can codify and implement non-formalized testing scripts that are still manually processed~\cite{dobslaw2020boundary,Haas2021ManualTesting} based on their natural language processing capabilities. For example, we can imagine a conversational testing agent interpreting and executing natural language testing guidelines written for humans, executing tools and seeking clarifications if needed via the middleware. As such non-formalized testing techniques or guidelines are intended for humans, they could be readily adopted as part of already-existing testing practices, which can improve developer acceptance of results~\cite{Winter2022bloomberg}, while also acting as explanations for any generated results~\cite{kang2023explainable}. At the greatest level of autonomy, LLMs would formulate and execute testing plans, while conversing with the developer on abstract terms. For example, in our example from the previous section, a human had to copy-and-paste the generated tests from an LLM and manually execute the tests; with the appropriate middleware, this process could be automated, and the developer would only need make higher-level decisions. As a result, this level of autonomy has the potential to significantly reduce the developer effort that goes into testing software. It could also lead to better utilisation of computer resources by continuously trying to fulfill testing goals even when the human\slash developer is away.

\section{Present-day limitations}
\label{sec:limitations}

% \rf{Check the 'many questions' slide of NEXTA keynote for more limitations and unknowns with this tech}

A major limitation to the attempt to use current generation LLMs for software testing in the way of \name is that on their own, LLMs lack any agency to use external tools. However, specific prompting techniques such as REACT~\cite{yao2022react} or PAL~\cite{gao2022pal} have shown that external tools can be indirectly woven into the dialogue, providing LLMs the information produced by external tools so that it can continue further inference using them. Also, systems like HuggingGPT~\cite{shen2023hugginggpt} and AutoGPT~\cite{autogpt2023} show that even if an LLM is not provided with tool access on a lower level it can be done via direct prompting and explanation.

A further limitation is that the planning abilities of current LLMs are not well-defined, often considered among their less developed competencies~\cite{bubeck2023sparks}. While this might be mitigated by multi-step prompting techniques as in recent work~\cite{yao2023tree}, other hybrid systems might need to be explored that combines LLMs with more traditional AI planning tools and algorithms.

%Another, indirect limitation might be the cost of training and operating these LLMs. The few-shot learning capabilities of LLMs has been associated with their sizes~\cite{kaplan2020scaling}, resulting in the rapid growth of model sizes~\cite{chowdhery2022palm,rae2022scaling}. Consequently, these models not only consume enormous energy to train and operate but also remain exclusive to organizations that are capable of providing massive amounts of computational resources, limiting generalizable tooling based on them. However, a recent study into the scaling behaviour suggests that the performance of these models are not singularly dependent on model sizes, but also on the volume of training data~\cite{hoffmann2022training}. Further, techniques such as model quantization~\cite{polino2018model} and low-rank adaptation~\cite{hu2021lora} have produced promising results in extracting and/or generating smaller language models that are capable of performances comparable to those of LLMs.

The significant costs of LLM training and operation constitute an indirect limitation, with e.g. their few-shot learning ability associated with their size~\cite{kaplan2020scaling}. 
%and associated size driving model growth. 
The ensuing model growth leads to significant energy requirements and limits access to resource-rich organizations, hence impeding development of open tools. Despite this, performance does not rely solely on size, but also on training data volume~\cite{hoffmann2022training}. Furthermore, techniques like model quantization and low-rank adaptation have shown promise in creating smaller, yet effective models~\cite{polino2018model, hu2021lora}, which due to their more permissive licenses can also mitigate some concerns about LLM use when dealing with confidential data.

\section{Conclusions}
\label{sec:conclusion}

This paper provides an overview of conversational and potentially autonomous testing agents by first presenting a taxonomy of such agents, describing how these agents could help developers (and increasingly so when granted with greater autonomy). A concrete example of a conversation with an LLM is provided as initial confirmation that conversational testing can be used to enhance the testing effectiveness of developers. Finally, limitations of these techniques is provided, providing context for our vision. As described in the paper, appropriate middleware is critical for realizing the autonomous testing agents that we envision; we plan on investigating which software engineering tools could aid the efficacy of conversational testing, and how they can be integrated harmoniously with LLMs to aid software testing in practice.

\section*{Acknowledgment}

Robert Feldt has been supported by the Swedish Scientific Council (No. 2020-05272, `Automated boundary testing for QUality of AI/ML modelS') and by WASP (`Software Boundary Specification Mining (BoundMiner)'). Sungmin Kang, Juyeon Yoon, and Shin Yoo were supported by the Institute for Information \& Communications Technology Promotion grant funded by the Korean government MSIT (No.2022-0-00995).

% \bibliographystyle{IEEEtran}
% \bibliography{IEEEabrv,llms}
\printbibliography

\balance{}

\end{document}